\documentclass[twocolumn,preprintnumbers,showpacs,amsmath,amssymb]{revtex4}
\usepackage{graphicx}
\usepackage{amsmath}
\usepackage{mathtools}
\usepackage{color}

\newcommand{\comm}[1]{}

\begin{document}

\title{ Meaning of delayed choice experiment and quantum uncertainty}
\author{Zinkoo Yun}
 \email{semiro@uvic.ca}
\affiliation{Department of Physics and Astronomy University of Victoria, Canada}

%\date{\ October 29, 2013}


\begin{abstract}
 By slight modifying of the delayed-choice experiment, it is argued that the quantum wave function must be interpreted as real physical entity; With this interpretation in mind, multiple least action paths due to uncertainty leads us to new perspective on the Compton wavelength and the uncertainty principle itself.  
\end{abstract}


\maketitle

\section{Introduction}
\begin{figure}
\begin{center}
    \includegraphics[height=7cm]{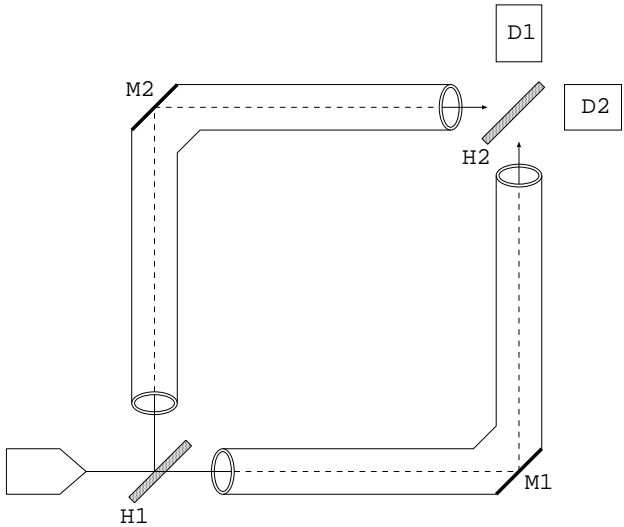}
  \end{center}
  \caption{Delayed-choice experiment. A stream of photons enters half-silvered mirror at H1. Without second half-silvered mirror at H2, one of detectors D1 and D2 clicks by 50/50 chances. As soon as we insert half-silvered mirror at H2, only one of detectors clicks.}
\label{delawh_fig1}
\end{figure}

Figure \ref{delawh_fig1} (without pipe wall) illustrates the delayed choice experiment proposed by John Wheeler\cite{wheeler}. We put a half-silvered mirror at H1. A stream of photons entering this half-silvered mirror show the probability 1/2 to penetrate it and the probability 1/2 to reflect it. The photon penetrating H1 reflects on full mirror at M1 and enters the detector D1. The photon reflecting H1 reflects on another full mirror at M2 and enters the detector D2. Without another half-silvered mirror at H2, any photon clicks either D1 or D2. It shows the particle property of photon;

If we put another half-silvered mirror at H2, and manipulated it properly, we observe only one of detectors always clicks, because after proper manipulating, two electromagnetic waves entering H2 interfere constructively at one side and interfere destructively at the other side. This shows the wave property of photon.

The most interesting event occurs when we insert half-silvered mirror at H2 after photon has passed H1 and M1/M2 but just before it reaches H2. We still observe only one of detector clicks. If you regard the photon traveled as a particle before inserting H2 (so it followed either path), it looks like this particle travels backward in time when it meets the half-silvered mirror at H2. It is widely accepted that the Delayed-choice experiment suggests that the choice of our experiment determines which property of duality to be displayed. 

 In section \ref{delchice_a} a slightly changed experiment is proposed to reveal the feature of wave function analogous to EPR pair. In section 3, multiple least action paths due to uncertainty are introduced and the dispersion speed of wave function compared to speed of light will lead us new perspective on the Compton wavelength. In section 4, a new interpretation about the uncertainty principle will be discussed.

\section{Interpretation of delayed-choice experiment}\label{delchice_a}
It is often quoted that this experiment shows that what is observed depends on the choice of experimental arrangement. It may better to say that this experiment demonstrates that the wave function is real physical entity which spreads on space time. To see this feature more clearly, separate two mirrors M1 and M2 far away each other (i.e., on two different galaxies) and enclose whole paths with pipe walls as shown in figure  \ref{delawh_fig1}. So that matter wave cannot affect each other all along the whole path except at H2. In other words, make sure two long paths are fully independent each other except at H2. 

A stream of photons pass through half-silvered mirror at H1. Without second half-silvered mirror H2, photon arrives either detector D1 or D2. Does it mean the photon followed one or the other path? As soon as we put a second half-silvered mirror H2, only one of detectors clicks. By adjusting inserting half-silvered mirror H2  a bit, we can make it in a way so that only the other detector clicks. It means that the photon followed both paths even before we insert half-silvered mirror H2.

This interpretation consistent with Feynman path integral point of view which says a particle follows all possible quantum paths simultaneously. Note that Feynman path integral is based on completeness of position quantum basis:
\begin{equation}\label{delchice1a} 
I=\sum_n \mid x_n\rangle\langle x_n\mid
\end{equation}

Before inserting H2, the wave function of a particle spreads in both paths and then collapse to one or the other detector when it reaches to them. We may say these two waves are entangled each other (like EPR pair) because they are two components of single wave function not the sum of two wave functions. Measuring one component collapses the other component. That is, if we measure the position eigenvalue (detector clicks) of one part, then immediately it collapse the other part to null state. This reminds us the measurement of EPR pair\cite{epr}.

There are experiments showing that single particle actually spreads in classically distinctive two positions. Using the superconducting quantum interference device (SQUID)\cite{friedman}, Friedman. {\it et al.} demonstrated that single particle actually travels two separate paths simultaneously.

The fact that we detect interference effect as soon as we insert H2 demonstrates that this wave function is objective real physical substance coming from both paths. From this, we understand that the Feynman path integral is not just mathematically formal description, but it describes real physical entity.\footnote{Someone may ask ``What do you mean by real?'' We may say the ``real'' is something  defined by fundamental physics laws. For example, Einstein field equations define energy momentum. In our case, Feynman path integral defines Feynman paths.} That is, a particle literally  follows all possible paths in the form of wave and reveals its particle property only when we measure the position eigen value. It means a particle is in general not local object (like a hard ball) and shows local property only when we measure its local property such as position eigen value. The only obstacle to accept this picture is our daily life experience. Simply because we are not used to it.

\section{Multiple least action paths}\label{delchice_b}
\begin{figure}
\begin{center}
    \includegraphics[height=6cm]{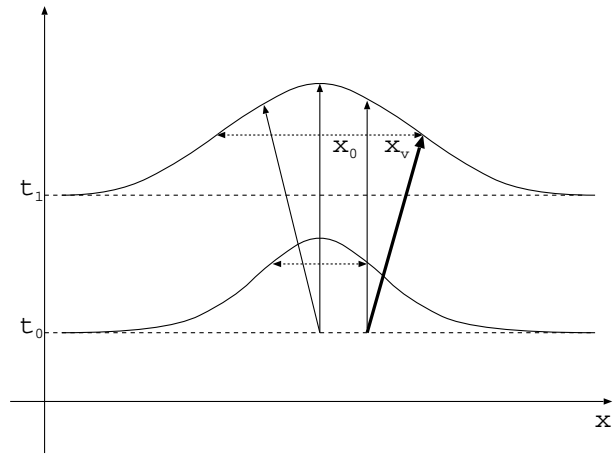}
  \end{center}
  \caption{In quantum mechanics due to uncertainty, there are many least action paths for given Lagrangian and initial condition.}
\label{delawh_fig2}
\end{figure}

In classical mechanics, the Lagrangian with the initial position and momentum determines the unique least action path. In quantum mechanics, due to the uncertainty relation between position and momentum, for given Lagrangian and initial wave function, there are many least action paths.  Figure \ref{delawh_fig2} shows dispersive evolution of static Gaussian wave packet representing static particle. We draw several least action paths by straight lines. 

From the initial uncertainty of position we can calculate final uncertainty. Suppose the initial uncertainty of position and momentum is $\Delta x_0$ and $\Delta v_0$ at $t=t_0$.
From the minimum uncertainty relation $m\Delta x_0 \Delta v_0=\hbar$, the uncertainty $\Delta x_v$ at $t_1$ due to the uncertainty $\Delta v_0$ is
\begin{equation}\label{delchice1b} 
\Delta x_v=\Delta v_0(t_1-t_0)=\frac{\hbar\Delta t_0}{m\Delta x_0}
\end{equation}
With the initial uncertainty $\Delta x_0$, the resulting position uncertainty $\Delta x_1$ at $t_1$ is
\begin{equation}\label{delchice1c} 
\Delta x_1=\sqrt{\Delta x^2_0+\Delta x^2_v}=\Delta x_0\sqrt{1+\Big(\frac{\hbar\Delta t_0}{m\Delta x^2_0}\Big)^2}
\end{equation}
We  can also derive (\ref{delchice1c})\cite{goswami} by putting a Gaussian wave packet in momentum space 
\begin{equation}\label{delchice1d} 
a(k)=\exp\Big[-\frac{(k-k_0)^2}{2(\Delta k)^2}\Big] 
\end{equation}
into that in position space
\begin{equation}\label{delchice1e} 
\psi(x,t)=\int^\infty_{-\infty} a(k,k_0)e^{i(kx-\omega t)}dk
\end{equation}
or by integrating Feynman kernels of all possible free paths.\cite{hibbs}

(\ref{delchice1c}) shows dispersive evolution of a static particle. Simple calculation tells that the dispersion speed of an electron initially within  its Compton wavelength is comparable to the speed of light while there is almost no dispersion within our lifetime for a mass of a pen with the same uncertainty. The reason is that because of its huge mass compare to an electron mass, a pen has very small $\Delta v_0$.

  This explains why in daily general life,  the object like a pen looks in exact position, and do not observe the effect of multiple position at the same time. In fact, no object is in position eigen state. It is always in multiple positions. As the mass is bigger, this multiple positions become so close each other enough to be indistinguishable classically. As the mass is bigger, the dispersion and the dispersion speed of the wave packet gets smaller, and we feel it is always in position eigenstate.   

\comm{
Note that the velocity uncertainty does not shrink as the position uncertainty grows.
For
\[
\Delta t_0=t_1-t_0,\quad \Delta t_1=t_2-t_1,\quad \Delta x'_v=\Delta v_0\Delta t_1=\frac{\hbar \Delta t_1}{m\Delta x_0}
\]
\[
\Delta x_2=\sqrt{\Delta x^2_1+\Delta x'^2_v}\cdots\textrm{I think this may not true.}
\]
\begin{eqnarray*}
\Delta x_2 &=& \sqrt{\Delta x^2_0+\Big(\frac{\hbar\Delta t_0}{m\Delta x_0}\Big)^2+ \Big(\frac{\hbar\Delta t_1}{m\Delta x_0}\Big)^2}\\
&\neq& \sqrt{\Delta x^2_0+\Big(\frac{\hbar(\Delta t_0+\Delta t_1)}{m\Delta x_0}\Big)^2}
\end{eqnarray*}
}

\section{Compton wavelength}\label{delchice_c}
In this section the fundamental meaning of the Compton wavelength will be revealed.
According to (\ref{delchice1c}), as the initial uncertainty $\Delta x_0$ gets smaller, the dispersion speed grows very large. As $\Delta x_0\to 0$, this speed can exceed the speed of light. It means if we confine the particle to very small region, after some moment $\Delta t$, we may discover it at the distance larger than $c\Delta t$. This violates the speed limit of relativity. Thus there must be minimum uncertainty $\Delta x_0$ corresponding to the mass of the particle.

The dispersive speed $v_{\textrm{disp}}$ is
\begin{equation}\label{delchice1f} 
v_{\textrm{disp}}=\frac{\Delta x_1-\Delta x_0}{\Delta t}=\frac{\Delta x_0}{\Delta t}\left(\sqrt{1+\Big(\frac{\hbar\Delta t}{m\Delta x^2_0}\Big)^2}-1\right)
\end{equation}
The dispersive speed grows as $\Delta x_0\to 0$ and it goes to zero as $\Delta t\to 0$.
For sufficiently large $\Delta t$, $v_{\textrm{disp}}$ can grow quite large. For sufficiently large $\Delta t$  and sufficiently small $\Delta x_0$,
\begin{equation}\label{delchice1g} 
v_{\textrm{disp}}\approx \frac{\Delta x_0}{\Delta t}\Big(\frac{\hbar\Delta t}{m \Delta x^2_0}\Big)=\frac{\hbar}{m\Delta x_0}
\end{equation}
\footnote{We can get (\ref{delchice1g}) simply from (\ref{delchice1b}) by $v_{\textrm{disp}}=\frac{\Delta x_v}{\Delta t_0}$.} 
which must be smaller than the speed of light.
\begin{equation}\label{delchice1h} 
\frac{\hbar}{m\Delta x_0}< c,\qquad \frac{\hbar}{mc}<\Delta x_0
\end{equation}
where $\frac{\hbar}{mc}$ is nothing but the Compton wavelength.

Thus the minimum uncertainty of the wave packet of a particle with mass $m$ cannot be smaller than its Compton wavelength. Thus, in principle, the eigen state of position for a particle does not exist. This is the fundamental physical meaning of the Compton wavelength.

For example, the Compton wavelength of proton is $\hbar/mc\sim 10^{-15} m$, and that of an electron is $\hbar/mc\sim 10^{-11} m$.
\comm{
For proton:
\begin{equation}\label{delchice1i} 
 \frac{\hbar}{mc}=\frac{6.6\cdot 10^{-34}m^2 kg/s}{1.6\cdot 10^{-27}kg\: 3\cdot 10^8 m/s}\sim 10^{-15} m
\end{equation}
For electron:
\begin{equation}\label{delchice1j} 
 \frac{\hbar}{mc}=\frac{6.6\cdot 10^{-34}m^2 kg/s}{9\cdot 10^{-31}kg\: 3\cdot 10^8 m/s}\sim 10^{-11} m
\end{equation}
}
Thus the uncertainty of electron is bigger than that of proton. The minimum uncertainty of electron is close to the size of proton. From the conclusion of section \ref{delchice_a} (The wave function itself is a physical entity.) we can understand that this is not just a coincidence. We can say that the size of an electron is much bigger than the size of proton. The lighter particle has bigger physical size which is counter intuitive. In contrast to traditional view, the space between proton and electron in hydrogen atom is not empty space. It is filled with single electron. This explains why the size of the hydrogen atom is inverse proportional to the mass of electron. 

This implies there should be no lighter elementary particle than electron in hydrogen atom. If there is, the size of hydrogen atom must be bigger.  We can understand that the electron must be the lightest elementary particle in most atoms.

Therefore the atom containing electrons cannot be smaller than $10^{-11}m$.
For example of hydrogen atom, the Bohr radius must be bigger than the Compton wavelength.
With $4\pi\epsilon_0=1$,
\begin{equation}\label{delchice1k} 
\frac{n^2\hbar^2}{m e^2} > \frac{\hbar}{mc}\to \frac{e^2}{n^2\hbar c}<1
\end{equation}
This requires the fine structure constant $\alpha\equiv \frac{e^2}{\hbar c}$  must be less than 1 to form a hydrogen atom with $n=1$. In other words, if the fine structure constant is bigger than 1, the hydrogen atom with $n=1$ cannot form.

Since $\hbar/mc$ is the minimum size of the object can have, any black hole also cannot collapse smaller than $\hbar/mc$. Even though this size is quite small for ordinary size of black holes, the classical point like singularity of a black hole does not exist in quantum world unless the mass of the black hole is infinity.

\comm{
In order for a particle to become a black hole, the minimum uncertainty of the particle $\frac{\hbar}{mc}$ must be smaller than its Schwarzchild radius, 
\begin{equation}\label{delchice1l} 
\frac{\hbar}{mc}<\frac{2Gm}{c^2}\to \sqrt{\frac{\hbar c}{2G}}< m \to 10^{-11}kg<m
\end{equation}
Thus the minimum mass of a black hole is $\sim 10^{-11}kg$.

We can calculate the corresponding maximum size of collapse (i.e., minimum mass) of black holes.
\begin{equation}\label{delchice1m}
\frac{2Gm}{c^2}=\frac{2G}{c^2}\sqrt{\frac{\hbar c}{2G}}=\sqrt{\frac{2G\hbar}{c^3}} 
\end{equation}
}

\section{Uncertainty principle}\label{delchice_d}
The concept of a particle as some form of point like object is just an illusion induced by  everyday life experience. It looks to us that everything around us is in position eigenstate. As $\langle x \mid\psi\rangle$ implies, observing position eigenvalue is just observing one of many quantum operators. If the state is in eigenstate of other operators, it is not fair to consider it is still point like object which we just uncertain its position. Because in its nature, it does not have exact position. The fundamental nature of particle is the wave function itself which itself is physical entity. The particle of conventional image exists nowhere or everywhere in the wave function. This reflects the lack of human word to describe something we never experience.

As we have seen in (\ref{delchice1h}), if the particle is localized smaller than its Compton wavelength, its dispersion velocity violates speed limit set by relativity. Thus a particle cannot localized at point-like eigen state of position. So the exact position of particle means incomplete information about the particle. The complete information about the particle is the wave function itself. If someone says he has the information about the exact position (+ exact momentum) of the particle, we know he is wrong. In fact, he has incomplete information about the particle because, in principle, there is no wave function of exact position eigenvalues as (\ref{delchice1h}) suggests. Thus he does not have complete information about the wave function, so his information about the particle is incomplete. His description is an approximation of wave function. Thus we understand that the uncertainty principle applies to classical mechanics not to quantum mechanics. Using assumed exact position (+ exact momentum) in calculation reflects how we are uncertain about the wave function and how we are uncertain about the system. 

A simple example which demonstrates the nature of uncertainty principle is the particle in a box:
\comm{
The probabilities $f(\vec{p})$ for each micro state corresponding $\vec{p}$ of a particle in a box with volume $L^3$ can be summed up to give a probability between $[p, p+dp]$:
\[
\sum_{ms}f(\vec{p}) \to \int^{\infty}_{-\infty}\frac{d^3p}{\Big(\frac{h}{L}\Big)^3}f(\vec{p})
\]
}
We can calculate the number of micro states of translational motion between $[p,p+dp]$  for a particle in a box with volume $L^3$ using boundary condition of quantum waves, $p_n=nh/L$:
\begin{equation}\label{mctn2}
\frac{d^3p}{\Big(\frac{h}{L}\Big)^3}
\end{equation}

On the other hand uncertainty principle in quantum mechanics states that
\begin{equation}\label{mctn4}
\Delta x \Delta y \Delta z \Delta p_x \Delta p_y \Delta p_z\geq h^3
\end{equation}
So there are $d^3x d^3p/h^3$ distinguishable states  of translational motion that a particle can occupy. 
In case of a particle in a box with volume $L^3$, it is $L^3 d^3p/h^3$ which is the same as (\ref{mctn2}) from boundary condition of quantum waves. 
\comm{
\begin{equation}\label{mctn3}
\sum_{ms}f(\vec{p}) \to \int \frac{d^3xd^3p}{h^3}f(\vec{p})
=\frac{L^3}{h^3}\int^\infty_{-\infty}d^3p f(\vec{p})
\end{equation}
}
 This implies that the uncertainty
principle is a direct result of boundary condition of quantum waves. In other words,
uncertainty in momentum is the interval between two momentum levels $p_{n+1}$ and $p_n$, \emph{not} 
limit on our accuracy in measuring momentum. From (\ref{mctn4}), the minimum uncertainty of momentum is,
\begin{equation}\label{mctn5}
L^3\Delta p^3\geq h^3, \qquad \Delta p\geq \frac{h}{L}
\end{equation}
which is  the boundary condition of quantum wave specifying the  the interval between $p_{n+1}$ and $p_n$. We cannot measure the momentum more accurately than $h/L$, because simply \emph{in principle} the micro states do not exist beyond that value.
%not because \emph{in practical} we have lack of ability of measuring devices beyond that.  This shows that a particle is a just one kind of property of a wave rather than a wave is a one kind of property of a particle or they are dual.
This shows there is no quantum uncertainty. Simply the state does not exist beyond certain value.
Claiming that someone has the information of exact momentum for a particle inside box exhibits the fact that he has uncertain  information about the particle.
\\

\section{Conclusion} 
By modifying the delayed choice experiment, we could understand that the wave function must be interpreted as real physical entity not just our knowledge about the system or just mathematically formal expression. 

With this property of wave function in mind, the dispersion of wave by multiple least action paths can be views as the dispersion of physical substance. In order not to violated the speed limit of relativity, this dispersion speed must be less than the speed of light. It requires that the dispersion of wave cannot be smaller than the Compton wavelength of it. Thus  we can consider the Compton wavelength as the size of the particle. Because of this property, the eigenstate of position more accurate than its Compton wavelength is physically impossible to exist. Thus the exact position in classical mechanics is just an approximate expression of wave function which is, in fact, the complete information about the system. Therefore the uncertainty principle applies to classical mechanics. There is no uncertainty principle in quantum mechanics.

%\section*{Acknowledgments}
%I would like to thank Werner Israel for his advises and useful comments on this work.

\end{document}